\documentclass[11pt]{article}
\usepackage{amssymb,latexsym,amsmath}
\usepackage[dvips]{graphicx}
\usepackage{cite}
\newtheorem{theo}{Theorem}

\newtheorem{prop}[theo]{Proposition}
\def\ii{\mathrm{i}} % imaginary unit i
\def\ee{\mathrm{e}} % exponential unit e
\def\su{\mathfrak{su}}
\def\uu{\mathfrak{u}}

\def\gl{\mathfrak{gl}}
\def\ssl{\mathfrak{sl}}
\def\osp{\mathfrak{osp}}
\def\diag{\mathop{\rm diag}\nolimits}
\def\str{\mathop{\rm str}\nolimits}
\def\qdots{\mathinner{\mkern1mu\raise1pt\vbox{\kern7pt\hbox{.}}\mkern2mu
 \raise4pt\hbox{.}\mkern2mu\raise7pt\hbox{.}\mkern1mu}}
\newcommand{\myatop}[2]{\genfrac{}{}{0pt}{}{#1}{#2}}
\def\mybox{\hfill$\Box$}
\begin{document}
\begin{center}
{\Large \bf
A finite oscillator model related to $\ssl(2|1)$}\\[5mm]
{\bf E.I.\ Jafarov\footnote{Permanent address: 
Institute of Physics, Azerbaijan National Academy of Sciences, Javid av.\ 33, AZ-1143 Baku, Azerbaijan}
and J.\ Van der Jeugt} \\[1mm]
Department of Applied Mathematics and Computer Science,
Ghent University,\\
Krijgslaan 281-S9, B-9000 Gent, Belgium\\[1mm]
E-mail: ejafarov@physics.ab.az, Joris.VanderJeugt@UGent.be
\end{center}

\vskip 10mm
\noindent
Short title: $\ssl(2|1)$ oscillator


%\addtolength{\baselineskip}{2mm}
%\addtolength{\abovedisplayskip}{1mm}
%\addtolength{\belowdisplayskip}{1mm}
%\addtolength{\parskip}{1mm}

\begin{abstract}
We investigate a new model for the finite one-dimensional quantum oscillator based upon the Lie superalgebra $\ssl(2|1)$.
In this setting, it is natural to present the position and momentum operators of the oscillator as odd elements
of the Lie superalgebra.
The model involves a parameter~$p$ ($0<p<1$) and an integer representation label~$j$.
In the $(2j+1)$-dimensional representations $W_j$ of $\ssl(2|1)$, the Hamiltonian has the usual equidistant
spectrum. The spectrum of the position operator is discrete and turns out to be of the form $\pm\sqrt{k}$, where
$k=0,1,\ldots,j$.
We construct the discrete position wave functions, which are given in terms of certain Krawtchouk polynomials.
These wave functions have appealing properties, as can already be seen from their plots.
The model is sufficiently simple, in the sense that the corresponding discrete Fourier transform
(relating position wave functions to momentum wave functions) can be constructed explicitly.
\end{abstract}

\section{Introduction}

Finite quantum harmonic oscillator models (simply referred to as finite oscillator models) are of importance
in optical image processing~\cite{Atak2005}, or in models where only a finite number of eigenmodes can exist such as in
signal processing~\cite{Atak1994,Atak1997,Atak1999b}. 
Quantum kinematics of finite oscillators has also been used to remove infinities or divergences in 
quantum theory or quantum field theory~\cite{Shiri2006}.
The main idea underlying these finite oscillator models is to replace the Heisenberg algebra of the standard
quantum oscillator, which allows only infinite-dimensional representations, by a ``closely related'' algebra with
the same dynamics but which allows finite-dimensional representations.

More concretely, for a one-dimensional finite oscillator one considers three (essentially self-adjoint) operators: 
a position operator $\hat q$, its corresponding momentum operator $\hat p$ and
a Hamiltonian $\hat H$ which is the generator of time evolution. 
These operators should satisfy the Hamilton-Lie equations (or the compatibility of Hamilton's equations with the Heisenberg
equations):
\begin{equation}
[\hat H, \hat q] = -\ii \hat p, \qquad [\hat H,\hat p] = \ii \hat q,
\label{Hqp}
\end{equation}
in units with mass and frequency both equal to~1, and $\hbar=1$.
The third relation of the canonical oscillator, $[\hat q, \hat p] = \ii$, is dropped since otherwise the corresponding
algebra (the Heisenberg algebra) has infinite-dimensional representations only. 
Instead, one requires~\cite{Atak2001}:
\begin{itemize}
\item all operators $\hat q$, $\hat p$, $\hat H$ belong to some (Lie) algebra (or superalgebra) $\cal A$;
\item the spectrum of $\hat H$ in (unitary) representations of $\cal A$ is equidistant.
\end{itemize}
The most popular model is with ${\cal A}= \su(2)$ (or its enveloping algebra), see~\cite{Atak2001,Atak2001b,Atak2005}.
This model is also underlying the theory in~\cite{Shiri2006}.
The relevant finite-dimensional representations are the common $\su(2)$ representations labelled 
by an integer or half-integer $j$.
In such a representation, the Hamiltonian is taken as $\hat H=J_0+j+\frac12$, where $J_0=J_z$ is the diagonal $\su(2)$
operator, and thus the spectrum of $\hat H$ is $n+\frac12$ ($n=0,1,\ldots,2j$).
The operators $\hat q$ and $\hat p$ are linear combinations of the other $\su(2)$ operators $J_+$ and $J_-$, such
that the relations~\eqref{Hqp} are satisfied. 
They turn out to have a finite spectrum given by $\{-j,-j+1,\ldots,+j\}$~\cite{Atak2001}, see Figure~1(a).
In this context, one can also construct the discrete position and momentum wave functions.
For the $\su(2)$ case, these are given by Krawtchouk functions (normalized
symmetric Krawtchouk polynomials), i.e.\ Krawtchouk polynomials with parameter $p=1/2$. 
These discrete wave functions have many interesting properties, and their shape is reminiscent of those of the canonical oscillator~\cite{Atak2001}. 

In~\cite{JSV2011,JSV2011b}, deformations of $\su(2)$ were considered by introducing a deformation parameter $\alpha>-1$.
For the even-dimensional representations~\cite{JSV2011} ($j$ half-integer), this led to an alternative model of
the finite oscillator with the spectrum of $\hat H$ again given by $n+\frac12$ ($n=0,1,\ldots,2j$),
and with the spectrum of the position operator $\hat q$ given by
\[
\pm(\alpha+k) \qquad (k=1,2,\ldots,j+\frac12),
\]
i.e.\ a finite and mostly equidistant spectrum apart from a gap of size $2\alpha+2$ in the middle,
see Figure~1(b). 
For the odd-dimensional representations~\cite{JSV2011b} ($j$ integer), this led to a similar model with the
same spectrum of $\hat H$ but with the spectrum of $\hat q$ given by (Figure~1(c))
\[
0, \quad \pm \sqrt{k(2\alpha+k+1)}, \qquad (k=1,\ldots,j).
\]
In both deformations, the position wave functions could be constructed explicitly and
turned out to be related to normalized Hahn (or dual Hahn) polynomials~\cite{JSV2011,JSV2011b}. 
It was also shown how these wave functions could be interpreted as the finite version of a paraboson oscillator.

Recall however that Wigner~\cite{Wigner} was among the first to drop the canonical commutation relation $[\hat q, \hat p] = \ii$,
proposing a model with relations~\eqref{Hqp} with the extra condition 
\begin{equation}
\hat H = \frac12(\hat p^2 + \hat q^2).
\label{H-W}
\end{equation}
This is known as the Wigner quantum oscillator (or paraboson oscillator)~\cite{Palev79,Palev82}.
This paraboson oscillator still has an infinite energy spectrum of the form $n+a$ (with a positive 
representation parameter $a$), 
and the spectrum of the position operator is the real axis. 
The canonical oscillator is recovered from the paraboson oscillator in the representation with $a=1/2$,
i.e.\ one of the representations of the paraboson oscillator coincides with the canonical 
oscillator~\cite{Mukunda,Ohnuki,JLV2008}.
The algebraic structure equivalent with~\eqref{Hqp} and~\eqref{H-W} is the Lie superalgebra 
$\osp(1|2)$~\cite{Ohnuki,JLV2008}.
In this context, the position and momentum operators $\hat q$ and $\hat p$ are {\em odd} elements of
the Lie superalgebra, whereas $\hat H$ is an {\em even} element.
This observation, and the fact that the canonical oscillator fits in one of the $\osp(1|2)$ representations,
leads to the idea that it is perhaps more natural to consider the position and momentum operators of
alternative oscillator models as {\em odd} elements of a Lie superalgebra rather than just (even) elements
of a Lie algebra.

Following this last idea, we propose and investigate in this paper a finite oscillator model based on 
the Lie superalgebra $\ssl(2|1)$.
Indeed, apart from $\osp(1|2)$, $\ssl(2|1)$ is the simplest Lie superalgebra that can be considered as
a superversion of the Lie algebra $\su(2)$~\cite{Scheunert1977}.
The Lie superalgebra $\ssl(2|1)$ has moreover a class of representations $W_j$ of dimension $2j+1$ ($j$ integer)
which are similar to the common $\su(2)$ representations, see Section~2.

In the finite oscillator model studied here in Section~3, the Hamiltonian $\hat H$ is a diagonal operator with
spectrum $n+\frac12$ ($n=0,1,\ldots,2j$) in $W_j$. 
The position operator $\hat q$ is an arbitrary (self-adjoint) odd element from $\ssl(2|1)$, and 
the form of the momentum operator $\hat p$ follows from~\eqref{Hqp};
the model introduces in a natural way a real parameter $p$ with $0<p<1$. 
Our investigation shows that the spectrum of $\hat q$ in the representation $W_j$ is still very simple,
given by
\[
\pm \sqrt{k} \qquad (k=0,1,\ldots,j),
\]
see Figure~1(d). 
In order to prove this statement, we construct the eigenvectors of $\hat q$ in $W_j$ explicitly.
For this construction, we need some properties of Krawtchouk polynomials (but now with general parameter~$p$),
a well known set of discrete orthogonal polynomials.

The properties of the position and momentum wave functions for the new $\ssl(2|1)$ oscillator model
are investigated in Section~4. 
In particular, we study first some plots of the wave functions for different values of the
parameter~$p$.
These discrete wave function plots are rather different from the $\su(2)$ plots in~\cite{Atak2001};
only for small $p$-values these plots show some similarity with the plots of the continuous canonical 
oscillator wave functions.
We explore also the behaviour of the discrete wave functions when the representation parameter $j$
tends to infinity, and discover a relation with the paraboson oscillator.

When both position wave functions and momentum wave functions are known explicitly, one can examine the
transformation that relates the two. In the canonical case this is just the Fourier transform.
In a finite oscillator model, this gives a discrete version of the Fourier transform, determined
by a Fourier matrix $F$.
For the $\ssl(2|1)$ model, we manage to give an explicit expression for an arbitrary
matrix element $F_{kl}$ of $F$, see Section~5.
The properties of the matrix $F$ are studied and prove to be similar to those of the standard Discrete Fourier Transform 
used in spectral analysis. 
Our paper closes with some concluding remarks in Section~6.

\section{The Lie superalgebra $\ssl(2|1)$ and a class of representations}

The Lie superalgebra $\gl(2|1)$ is well known: it can be defined as the set of all
$3\times 3$ matrices $x=(x_{ij})$ with rows and columns labelled by indices $i,j=1,2,3$. 
As a basis in $\gl(2|1)$ we choose the Weyl matrices $e_{ij}, \; i,j=1,2,3$, where 
the odd elements are $\{e_{1i}, e_{i1} | i=2,3\}$,
and the remaining elements are even.
The Lie superalgebra bracket is determined by
\begin{equation}
[\![ e_{ij}, e_{kl} ]\!] \equiv e_{ij} e_{kl}-(-1)^{{\deg(e_{ij})
\deg(e_{kl})}}e_{kl}e_{ij} = \delta_{jk} e_{il} - (-1)^{\deg(e_{ij}) \deg(e_{kl})}
\delta_{il} e_{kj}. \label{Weyl}
\end{equation}

Note that the even part $\gl(2|1)_0 = \gl(2) \oplus \gl(1)$, where $\gl(2)=\hbox{span}\{ e_{ij}|i,j=1,2\}$
and $\gl(1) =\hbox{span}\{ e_{33}\}$. For elements $x$ of
$\gl(2|1)$, one defines the supertrace
as $\str(x)=x_{11}+x_{22}-x_{33}$. The Lie superalgebra $\gl(2|1)$ is
not simple, and one can define the simple superalgebra $\ssl(2|1)$
as the subalgebra consisting of elements with supertrace $0$. 
As a basis for $\ssl(2|1)$, it is convenient to follow the choice of~\cite{Frappat}, where one
can find the actual matrices of the basis~\cite[p.~261]{Frappat}:
\begin{align}
& F^+=e_{32},\ G^+=e_{13},\ F^-=e_{31},\ G^-=e_{23}, \label{odd-el}\\
& H=\frac12(e_{11}-e_{22}),\ E^+=e_{12},\ E^-=e_{21},\ Z=\frac12(e_{11}+e_{22})+e_{33}. \label{even-el}
\end{align}
So $\ssl(2|1)$ has four odd (or `fermionic') basis elements $F^+, F^-, G^+, G^-$ and four even
(or `bosonic') basis elements $H, E^+, E^-, Z$. The basis for the $\gl(2)$ subalgebra is $\{H,E^+,E^-\}$ and 
the $\gl(1)\cong U(1)$ subalgebra is spanned by $Z$.
The basic Lie superalgebra brackets can be found in~\cite[p.~261]{Frappat} or, in a different notation, in~\cite{Scheunert1977}.
For the odd elements, the anti-commutators are given by
\begin{align}
& \{F^\pm,G^\pm\}=E^\pm, \quad \{F^\pm,G^\mp\}=Z\mp H; \nonumber\\
& \{F^\pm,F^\pm\}=\{G^\pm,G^\pm\}=\{F^\pm,F^\mp\}=\{G^\pm,G^\mp\}=0. \label{co-odd}
\end{align}
For the even elements, the commutators are given by
\begin{equation}
[H,E^\pm]=\pm E^\pm,\quad [E^+,E^-]=2H,\quad [Z,H]=[Z,E^\pm]=0. \label{co-even}
\end{equation}
The mixed commutation relations read:
\begin{align}
&[H,F^\pm]=\pm\frac12 F^\pm, \quad [Z,F^\pm]=\frac12 F^\pm, \quad [E^\pm,F^\pm]=0,\quad [E^\mp,F^\pm]=-F^\mp; \nonumber\\
&[H,G^\pm]=\pm\frac12 G^\pm, \quad [Z,G^\pm]=-\frac12 G^\pm, \quad [E^\pm,G^\pm]=0,\quad [E^\mp,G^\pm]=G^\mp. \label{co-mixed}
\end{align}

The irreducible representations of $\ssl(2|1)$ have been studied by Scheunert {\em et al}~\cite{Scheunert1977} and Marcu~\cite{Marcu};
for a summary, see~\cite[\S 2.53]{Frappat}. 
The superalgebra $\ssl(2|1)$ has typical and atypical irreducible representations.
Here, we shall consider a class of atypical irreducible representations, labelled by a non-negative integer $j$
(these are denoted by $\pi_-(j/2)$ in~\cite{Frappat}).
In order to describe these representations explicitly, let us choose a basis for the representation space $W_j$
of the form
\begin{equation}
|j,m\rangle, \qquad m=-j,-j+1,\ldots,+j.
\end{equation}
So $\dim(W_j)=2j+1$. For the actions of the $\ssl(2|1)$ basis elements on these vectors, it is handy to
use the following ``even'' and ``odd'' functions, defined on integers $n$:
\begin{align}
& {\cal E}(n)=1 \hbox{ if } n \hbox{ is even and 0 otherwise},\nonumber\\
& {\cal O}(n)=1 \hbox{ if } n \hbox{ is odd and 0 otherwise}. \label{EO}
\end{align}
Note that ${\cal O}(n)=1-{\cal E}(n)$, but it is convenient to use both notations.
The actions of the odd generators are now given by:
\begin{align}
& F^\pm |j,m\rangle = \pm {\cal O}(j-m) \sqrt{\frac{j\pm m+1}{2}}\; |j,m\pm 1\rangle, \nonumber\\
& G^\pm |j,m\rangle = \pm {\cal E}(j-m) \sqrt{\frac{j\mp m}{2}} \;|j,m \pm 1\rangle.
\label{FGact}
\end{align}
The actions of the even generators can in principle be computed from~\eqref{co-odd}, and are
\begin{align}
& Z |j,m\rangle = -{\cal E}(j-m) \frac{j}{2}\; |j,m\rangle -{\cal O}(j-m) (\frac{j+1}{2})\; |j,m\rangle , \nonumber\\
& H |j,m\rangle = \frac{m}{2}\; |j,m\rangle , \nonumber\\
& E^\pm |j,m\rangle = \frac12{\cal E}(j- m) \sqrt{(j\mp m)(j\pm m+2)}\; |j,m\pm 2\rangle \nonumber\\
& \qquad +\frac12{\cal O}(j-m) \sqrt{(j \mp m-1)(j\pm m+1)}\; |j,m\pm 2\rangle.
\label{EHZact} 
\end{align}
It is easy to verify that the relations~\eqref{co-odd}-\eqref{co-mixed} are satisfied for these actions.
One can also see that with respect to the even subalgebra $\gl(2)\oplus \gl(1) \cong \su(2) \oplus U(1)$, 
$W_j$ decomposes as $(\frac{j}{2}; -\frac{j}{2}) \oplus (\frac{j-1}{2}; -\frac{j+1}{2})$, where $(l;b)$ denotes the
$\su(2) \oplus U(1)$ representation ``with isospin $l$ and hypercharge $b$'' \cite{Frappat}.

The above representation is a star representation (or unitary representation) for the adjoint operation
\begin{equation}
Z^\dagger=Z,\quad H^\dagger=H,\quad (E^\pm)^\dagger = E^\mp,\quad
(F^\pm)^\dagger = -G^\mp,\quad (G^\pm)^\dagger = -F^\mp,
\label{adjoint}
\end{equation}
compatible with the positive definite inner product on the representation space $W_j$:
\begin{equation}
\langle j,m | j, m'\rangle = \delta_{m,m'}.
\end{equation}

\section{The $\ssl(2|1)$ model for a finite one-dimensional oscillator}

We wish to investigate how $\ssl(2|1)$ can be used for a finite oscillator model. 
Inspired by the seminal paper~\cite{Atak2005} on the $\su(2)$ model for a finite oscillator, and by
the requirements for the spectrum of the Hamiltonian $\hat H$, one should take 
\begin{equation}
\hat H = 2H +j+\frac12
\end{equation}
as operator for $\hat H$ in the representation space $W_j$. This operator is diagonal, self-adjoint, and has indeed 
the equidistant spectrum:
$n+\frac12$ ($n=0,1,2,\ldots, 2j$).
Next comes the choice for the position operator $\hat q$, which should be a self-adjoint element of $\ssl(2|1)$.
There are two reasons to choose an odd element of $\ssl(2|1)$. 
First, choosing an even element would yield a model that is essentially the same as the $\su(2)$ model of~\cite{Atak2005}.
Second, as explained in the introduction, in Wigner's paraboson oscillator the position and momentum operator are 
elements of the odd part of the Lie superalgebra $\osp(1|2)$~\cite{Wigner,Ohnuki,JLV2008}. 
And since the canonical oscillator is a special case of the paraboson oscillator (corresponding to one particular 
$\osp(1|2)$ representation~\cite{Ohnuki,JLV2008}), the position and momentum operator
can also be considered as odd elements of a superalgebra in a particular representation.
The most general real self-adjoint odd element of $\ssl(2|1)$ is given by
\begin{equation}
A F^+ +B G^+-B F^- -AG^-,
\label{AFBG}
\end{equation}
with $A$ and $B$ real constants. 
An overall constant does not play a crucial role, so let us assume that there is some normalization like $A^2+B^2=1$. 
Consider now first the case that $A$ and $B$ have the same sign, say both positive
(the case $A$ positive and $B$ negative will be very similar, and is described at the end of this section).
In that case, one can write
\begin{equation}
\hat q = \sqrt{p}\; F^+ + \sqrt{1-p}\; G^+-\sqrt{1-p}\; F^- -\sqrt{p}\;G^-, \qquad (0 \leq p \leq 1).
\label{hatq}
\end{equation}
We shall in fact consider $0<p<1$ and later view the values $p=0$ and $p=1$ as a limit.

Once $\hat q$ is fixed, the form of $\hat p$ follows from the first equation of~\eqref{Hqp}, and thus
\begin{equation}
\hat p = \ii (\sqrt{p}\; F^+ + \sqrt{1-p}\; G^+ +\sqrt{1-p}\; F^- +\sqrt{p}\;G^-), \qquad (0 \leq p \leq 1).
\label{hatp}
\end{equation}
With these operators, \eqref{Hqp} and the conditions described in Section~1 are satisfied, and we can truly speak of
an $\ssl(2|1)$ model for the oscillator.

Now it remains to study the operators $\hat q$ and $\hat p$, in particular their spectrum and eigenvectors in 
the representation $W_j$.
Note that, due to the actions~\eqref{FGact}, in the (ordered) basis 
$\{ |j,j\rangle, |j,j-1\rangle, \ldots, |j,-j+1\rangle, |j,-j\rangle \}$ of $W_j$,
the operators $\hat q$ and $\hat p$ are tridiagonal matrices. In particular, for $\hat q$, one has:
\begin{equation}
\hat q=\left(
\begin{array}{cccccccc}
0 & R_1& 0 & \cdots & & & & 0 \\
R_1 & 0 & S_1 & \cdots & & & & 0\\
0 & S_1 & 0 & R_2 & & & & \vdots  \\
\vdots & & R_2 & 0 & S_2 & & & \\
 & & & S_2 & 0 & \ddots & &\vdots \\
 & & & & \ddots & \ddots & R_j & 0 \\
 & & & & & R_j & 0 & S_j \\
0 & \cdots & & & \cdots & 0 & S_j & 0 
\end{array}
\right) \equiv M_q,
\label{Mq} 
\end{equation}
where
\begin{equation}
R_k=\sqrt{p}\sqrt{j+1-k}, \qquad S_k=\sqrt{1-p}\sqrt{k} \qquad (k=1,2,\ldots,j).
\label{RS}
\end{equation}
The matrix form $M_p$ of $\hat p$ is similar.
For these matrices, we need to study the spectrum and the eigenvectors.
It is at this point that Krawtchouk polynomials play a role. 
Krawtchouk polynomials $K_n(x;p, N)$ of degree $n$ in the variable $x$, 
with parameter $p$ are defined by~\cite{Koekoek,Ismail,Andrews}:
\begin{equation}
K_n(x;p,N) = {\;}_2F_1 \left( \myatop{-n,-x}{-N} ; \frac{1}{p} \right), \qquad (n=0,1,\ldots,N)
\label{defK}
\end{equation}
in terms of the hypergeometric series $_2F_1$~\cite{Andrews,Bailey,Slater}
(which is terminating here because of the appearance of $-n$ as numerator parameter).
Their (discrete) orthogonality relation reads~\cite{Koekoek,Ismail,Andrews}:
\begin{equation}
\sum_{x=0}^N w(x;p,N) K_l(x;p, N) K_n(x;p,N) = h(n;p,N)\, \delta_{ln}, \qquad (0<p<1)
\label{orth-K}
\end{equation} 
where
\begin{equation}
w(x;p,N) = \binom{N}{x} p^x(1-p)^{N-x}  \quad (x=0,1,\ldots,N); \qquad
 h(n;p,N)= \frac{n!(N-n)!}{N!} \left( \frac{1-p}{p}\right)^n.
\end{equation}
For the orthonormal Krawtchouk functions we use the notation:
\begin{equation}
\tilde K_n(x;p,N) \equiv \frac{\sqrt{w(x;p,N)}}{\sqrt{h(n;p,N)}}\, K_n(x;p,N).
\label{K-tilde}
\end{equation}
Now we are able to describe the eigenvalues and (orthonormal) eigenvectors of $M_q$.
In this context, two sets of Krawtchouk polynomials play a role: $K_n(x;p,j)$ (with $N=j$) and 
$K_n(x;p,j-1)$ (with $N=j-1$).
\begin{prop}
Let $M_q$ (i.e.\ the matrix representation of $\hat q$) be the tridiagonal $(2j+1)\times(2j+1)$-matrix~\eqref{Mq} 
and let $U=(U_{kl})_{0\leq k,l\leq 2j}$
be the $(2j+1)\times(2j+1)$-matrix with matrix elements:
\begin{align}
& U_{2n,j} =(-1)^n \tilde K_0(n;p,j), \;n\in\{0,1,\ldots,j\};\;
U_{2n+1,j} = 0, \; n\in\{0,\ldots,j-1\}; \label{Uj}\\
& U_{2n,j-k} = U_{2n,j+k} = \frac{(-1)^n}{\sqrt{2}} \tilde K_k(n;p,j)
, \;n\in\{0,1,\ldots,j\}, \; k\in\{1,\ldots,j\};  \label{Ueven}\\
& U_{2n+1,j-k} = -U_{2n+1,j+k} = -\frac{(-1)^n}{\sqrt{2}} \tilde K_{k-1}(n;p,j-1), \; 
n\in\{0,1,\ldots,j-1\}, \quad k \in\{1,\ldots,j\}\label{Uodd}.
\end{align}
Then $U$ is an orthogonal matrix:
\begin{equation}
U U^T = U^TU=I.
\end{equation}
The columns of $U$ are the eigenvectors of $M_q$, i.e.
\begin{equation}
M_q U = U D,
\label{MUUD}
\end{equation}
where $D$ is a diagonal matrix 
containing the eigenvalues of $M_q$:
\begin{equation}
D=\diag(-\sqrt{j},-\sqrt{j-1},\ldots,-\sqrt{2}, -1,0,1,\sqrt{2}, \ldots,\sqrt{j-1},\sqrt{j}). \label{D} 
\end{equation}
\label{propU}
\end{prop}

\noindent {\bf Proof.}
Using the orthogonality of the Krawtchouk polynomials, and the explicit expressions~\eqref{Uj}-\eqref{Uodd}, 
a simple computation shows that $(U^TU)_{kl}=\delta_{kl}$. Thus $U^TU=I$, the identity matrix, and 
hence $UU^T=I$ holds as well.

It remains to verify~\eqref{MUUD} and that the eigenvalues are indeed~\eqref{D}. Due to the tridiagonal 
form~\eqref{Mq} of $M_q$, one has:
\begin{align}
&\big(M_qU\big)_{2n,k}= S_{n}U_{2n-1,k}+R_{n+1}U_{2n+1,k}, \label{MUeven}\\
&\big(M_qU\big)_{2n+1,k}= R_{n+1}U_{2n,k}+S_{n+1}U_{2n+2,k}.\label{MUodd}
\end{align}
For the first case~\eqref{MUeven}, we need to consider three distinct subcases, according to 
$k$ belonging to $\{0,1,\ldots,j-1\}$, to $\{j+1,j+2,\ldots,2j\}$ or $k=j$. 
For $k\in\{0,1,\ldots,j-1\}$, this gives:
\begin{align*}
&(M_qU)_{2n,j-k}=S_{n}U_{2n-1,j-k}+R_{n+1}U_{2n+1,j-k}  \\
& =\frac{(-1)^{n}}{\sqrt{2}}\sqrt{1-p}\sqrt{n}\tilde K_{k-1}(n-1;p,j-1) +
\frac{(-1)^{n+1}}{\sqrt{2}}\sqrt{p}\sqrt{j-n}\tilde K_{k-1}(n;p,j-1) \\
&=\frac{(-1)^{n+1}}{\sqrt{2}} 
\frac{(\sqrt{p})^{n+1} (\sqrt{1-p})^{j-n-1} \sqrt{(j-1)!}}{\sqrt{n!(j-n)!h(k-1;p,j-1)}} \\
& \qquad \times [ (j-n) K_{k-1}(n;p,j-1)- n(\frac{1-p}{p})K_{k-1}(n-1;p,j-1)].
\end{align*}
For the last linear combination between squared brackets, the backward shift operator formula
for Krawtchouk polynomials~\cite[(9.11.8)]{Koekoek} can be applied, and yields
\begin{align*}
(M_qU)_{2n,j-k}& =\frac{(-1)^{n+1}}{\sqrt{2}} 
\frac{(\sqrt{p})^{n+1} (\sqrt{1-p})^{j-n-1} \sqrt{(j-1)!}}{\sqrt{n!(j-n)!h(k-1;p,j-1)}}
j K_k(n;p,j) \\
&= -\sqrt{k}\; U_{2n,j-k} = \big(UD\big)_{2n,j-k}.
\end{align*}
For the other two subcases with first index $2n$, the computations are similar.
For the case~\eqref{MUodd}, we need again to consider three subcases.
Now one finds for $k\in\{0,1,\ldots,j-1\}$:
\begin{align*}
&(M_qU)_{2n+1,j-k}=R_{n+1}U_{2n,j-k}+S_{n+1}U_{2n+2,j-k}  \\
& =\frac{(-1)^{n}}{\sqrt{2}}\sqrt{p}\sqrt{j-n}\tilde K_{k}(n;p,j) +
\frac{(-1)^{n+1}}{\sqrt{2}}\sqrt{1-p}\sqrt{n+1}\tilde K_{k}(n+1;p,j) \\
&=\frac{(-1)^{n+1}}{\sqrt{2}} 
\frac{(\sqrt{p})^{n+1} (\sqrt{1-p})^{j-n} \sqrt{j!}}{\sqrt{n!(j-n-1)!h(k;p,j)}}
[ K_{k}(n+1;p,j)- K_{k}(n;p,j)].
\end{align*}
For the last linear combination, the forward shift operator formula
for Krawtchouk polynomials~\cite[(9.11.6)]{Koekoek} can be applied, and yields
\begin{align*}
(M_qU)_{2n+1,j-k}& =
- \frac{(-1)^{n+1}}{\sqrt{2}} 
\frac{(\sqrt{p})^{n+1} (\sqrt{1-p})^{j-n} \sqrt{j!}}{\sqrt{n!(j-n-1)!h(k;p,j)}} \frac{k}{pj} K_{k-1}(n;p,j-1) \\
&= -\sqrt{k}\; U_{2n+1,j-k} = \big(UD\big)_{2n+1,j-k}.
\end{align*}
The other two subcases are similar.
This completes the proof.
\mybox
  
The above proposition gives, besides the eigenvectors, also the spectrum of the position operator $\hat q$
in the representation $W_j$. 
It will be appropriate to denote these $\hat q$-eigenvalues by $q_k$, where $k=-j,-j+1,\ldots,+j$, so
\begin{equation}
q_{\pm k} = \pm\sqrt{k}, \qquad k=0, 1, \ldots, j.
\label{spec-q}
\end{equation}
To our knowledge, this is the first time we come across a tridiagonal operator with such a spectrum.
Note also that the spectrum of $\hat q$ is independent of the parameter $p$ in~\eqref{hatq}, but the eigenvectors
themselves do depend on $p$.

As far as the eigenvectors of $\hat q$ are concerned, these are the columns of the matrix $U$. 
It will be useful to introduce a notation for these eigenvectors:
the orthonormal eigenvector of the position operator $\hat q$ in $W_j$ for the eigenvalue $q_k$, denoted by $|j,q_k)$, is given 
in terms of the standard basis by
\begin{equation}
|j,q_k) = \sum_{m=-j}^j U_{j+m,j+k} |j,-m\rangle.
\end{equation}

For the matrix representation $M_p$ of $\hat p$, the analysis is essentially the same. 
Without going into the details, we give the final result:
\begin{prop}
Let $M_p$ be the tridiagonal $(2j+1)\times(2j+1)$-matrix representing $\hat p$ in $W_j$, 
and let $V=(V_{kl})_{0\leq k,l\leq 2j}$
be the $(2j+1)\times(2j+1)$-matrix with matrix elements
\begin{equation}
V_{2k,l}=-\ii (-1)^k U_{2k,l},\qquad V_{2k+1,l}= (-1)^k U_{2k+1,l},
\label{V}
\end{equation}
where $U$ is the matrix determined by \eqref{Uj}-\eqref{Uodd}.
Then $V$ is a unitary matrix, $V V^\dagger = V^\dagger V=I$.
The columns of $V$ are the eigenvectors of $M_p$, i.e.
\begin{equation}
M_p V = V D,
\label{MVVD}
\end{equation}
where $D$ is the same diagonal matrix as in Proposition~1. In other words, the eigenvalues of $\hat p$
are also given by:
\begin{equation}
-\sqrt{j},-\sqrt{j-1},\ldots,-\sqrt{2}, -1,0,1,\sqrt{2}, \ldots,\sqrt{j-1},\sqrt{j}.
\end{equation}
The matrix $V$ of eigenvectors satisfies:
\begin{equation}
V^T V = \left( \begin{array}{cccc}
0 & \cdots & 0 & -1 \\ 0 & \cdots & -1 & 0 \\ \vdots & \qdots & \vdots &\vdots\\ -1 & \cdots & 0 & 0 \\ \end{array}\right),
\label{VTV}
\end{equation}
and
\begin{equation}
V={\cal J}U \qquad\hbox{where}\qquad
{\cal J}=-\ii \diag(\ii^0,\ii^1,\ii^2,\ii^3,\ldots, \ii^{2j}) = \diag(-\ii,1,\ii,-1,\ldots).
\label{VJU}
\end{equation}
\end{prop}
The last assertions follow from the explicit
expressions~\eqref{V}, \eqref{Uj}-\eqref{Uodd} and the orthogonality properties of Krawtchouk polynomials.

Also here, it will be useful to denote the $\hat p$-eigenvalues by $p_k$, where $k=-j,-j+1,\ldots,+j$
(so $p_{\pm k} = \pm\sqrt{k}$, $k=0, 1, \ldots, j$), and to write the normalized eigenvectors as:
\begin{equation}
|j,p_k) = \sum_{m=-j}^j V_{j+m,j+k} |j,-m\rangle.
\end{equation}

We end this section with two remarks. 
First, let us briefly return to the remaining case~\eqref{AFBG} with $A$ positive and $B$ negative;
or, without losing generality, $A=\sqrt{p}$ and $B=-\sqrt{1-p}$.
The corresponding matrix $M_q'$ is then the same as in~\eqref{Mq}-\eqref{RS}, but with $S_k$ replaced by $-\sqrt{1-p}\sqrt{k}$.
In other words, one can write $M_q'=D_1 M_q D_1$ where $D_1=\diag(1,1,-1,-1,1,1,-1,-1,1,1,\ldots)$. 
This implies that $M_q' U'= U' D$, where $D$ is the same matrix as in Proposition~1 and $U'=D_1U$.
To conclude for this second case: the eigenvalues remain the same, and the matrix $U'$ of eigenvectors is the same 
as that of the first case, up to sign changes in rows.
For this reason, we shall not return to this second case, and just continue with the first case~\eqref{hatq}
for our analysis.

Secondly, we have so far considered~\eqref{hatq} with $0<p<1$, but what about the cases $p=0$ and $p=1$? 
In these limiting cases, the form of $M_q$ in~\eqref{Mq} remains valid, and since the eigenvalues of $M_q$ in~\eqref{D}
are independent of $p$, the eigenvalues are again given by~\eqref{D}. For the matrix of eigenvectors $U$, one
can simply take the right limit $p\rightarrow 0$ or the left limit $p\rightarrow 1$ in the matrix $U$
given in Proposition~1. For example, it is easy to compute:
\begin{equation}
U_0=\lim_{ \myatop{\scriptstyle p\rightarrow 0}{\scriptstyle p>0}} U, \qquad
U_0 = \frac{1}{\sqrt{2}} 
 \left( \begin{array}{ccccccc}
 \cdots & 0 & 0 & \sqrt{2} & 0 & 0 & \cdots \\
        & 0 & -1 & 0 & 1 & 0 & \\
        & 0 & 1 & 0 & 1 & 0 & \\
        & -1 & 0 & 0 & 0 & 1 & \\
 \cdots & 1 & 0 & 0 & 0 & 1 &  \cdots\\
        & \vdots &&&& \vdots &   
 \end{array}\right),
\label{U0}
\end{equation}     
where the general form of the orthogonal matrix $U_0$ (with only two nonzero elements per row, starting
from the second row onwards) is clear from the above. 
The left limit $p\rightarrow 1$ yields a similar matrix form for $U$.

\section{Position and momentum wave functions and their properties}

The position (resp.\ momentum) wave functions of the $\ssl(2|1)$ finite oscillator are the overlaps 
between the $\hat q$-eigenvectors (resp.\ $\hat p$-eigenvectors)
and the $\hat H$-eigenvectors (or equivalently, the $J_0$-eigenvectors $|j,m\rangle$).
Let us denote them by $\phi^{(p)}_{j+m}(q)$ (resp.\ $\psi^{(p)}_{j+m}(\bar p)$ ), where $m=j,j-1,\ldots,-j$, and where $q$ (resp.\ $\bar p$) assumes one of the 
discrete values $q_k$ (resp.\ $p_k$) $(k=-j,-j+1,\ldots,+j)$. 
Observe that we have denoted the momentum variable of the wave function by $\bar p$ in order not to 
confuse with the parameter $p$ ($0<p<1$) which appears in expression~\eqref{hatq} and which is also
the parameter of the Krawtchouk polynomials occurring here.
In the notation of the previous section (and where we want to emphasize the dependence on the parameter $p$), we have
\begin{align}
& \phi^{(p)}_{j+m}(q_k)= \langle j,-m | j,q_k ) = U_{j+m,j+k}, \label{Uphi}\\
& \psi^{(p)}_{j+m}(p_k)= \langle j,-m | j,p_k ) = V_{j+m,j+k}. \label{Vpsi}
\end{align}
Let us consider the explicit form of these wave functions, first for the position variable.
For $j+m$ even, $j+m=2n$, and for positive $q$-values one has
\begin{equation}
\phi^{(p)}_{2n} (q_{k}) = \frac{(-1)^n}{\sqrt{2}} \tilde K_{k}(n;p,j), \qquad n=0, 1, \ldots, j, 
\qquad k=1, \ldots, j,
\end{equation}
or equivalently, with $q_k=\sqrt{k}$ ($k=1,2,\ldots,j$):
\begin{equation}
\phi^{(p)}_{2n} (q_{k}) = \frac{(-1)^n}{\sqrt{2}} j!
\sqrt{\frac{p^{n+k}(1-p)^{j-n-k}}{n!(j-n)!k!(j-k)!}} 
{\ }_2F_1 \left( \myatop{-k,-n}{-j} ; \frac{1}{p} \right).
\label{phi-even}
\end{equation}
The expression for $\phi^{(p)}_{2n} (q_{-k})$, with $q_{-k}=-\sqrt{k}$ ($k=1,2,\ldots,j$) is also given by
the right hand side of~\eqref{phi-even}. So $\phi^{(p)}_{2n}$ is an even function. For the argument $0$, one
simply has
\begin{equation}
\phi^{(p)}_{2n} (0) = (-1)^n \sqrt{\binom{j}{n}p^{n}(1-p)^{j-n}} .
\label{phi-0}
\end{equation}

When $j+m$ is odd, $j+m=2n+1$, one finds for positive $q$-values
\begin{equation}
\phi^{(p)}_{2n+1} (q_{k}) = \frac{(-1)^n}{\sqrt{2}} \tilde K_{k-1}(n;p,j-1), \qquad n=0, 1, \ldots, j-1, 
\qquad k=1, \ldots, j,
\end{equation}
or, with $q_k=\sqrt{k}$ ($k=1,2,\ldots,j$):
\begin{equation}
\phi^{(p)}_{2n+1} (q_{k}) = \frac{(-1)^n}{\sqrt{2}} (j-1)!
\sqrt{\frac{p^{n+k-1}(1-p)^{j-n-k}}{n!(j-1-n)!(k-1)!(j-k)!}} 
{\ }_2F_1 \left( \myatop{-k+1,-n}{-j+1} ; \frac{1}{p} \right).
\label{phi-odd}
\end{equation}
The expression for $\phi^{(p)}_{2n+1} (q_{-k})$, with $q_{-k}=-\sqrt{k}$ ($k=1,2,\ldots,j$) is given by minus
the right hand side of~\eqref{phi-odd}, and $\phi^{(p)}_{2n+1} (0)=0$. So $\phi^{(p)}_{2n+1}$ is an odd function. 

Let us now consider some plots of these discrete wave functions. 
In Figure~2 we have plotted these functions for the representation $j=10$ (so discrete plots with $2j+1=21$ points).
We have considered three values for the parameter $p$: $p=0.1$, $p=1/2$ and $p=0.9$. 
The spectrum of $\hat q$ is independent of $p$, so for each of these three cases we plot points 
corresponding to the values $\pm\sqrt{k}$ ($k=0,1,\ldots,10$) on the horizontal axis.
For each of the considered $p$-values, we have plotted $\phi^{(p)}_0$ (the ground state), $\phi^{(p)}_1$ (the
first excited state), $\phi^{(p)}_2$ and $\phi^{(p)}_3$.

The behaviour of these discrete wave functions is reminiscent of that of the corresponding continuous
wave functions of the canonical oscillator, especially when $p$ is small. 
As $p$ increases, the wave function values for positions near the origin tend to decrease. 
In fact, for increasing $p$-values, the behaviour of the discrete wave functions rather tends
to the corresponding wave functions of the paraboson oscillator (see e.g.\ Figure~3 of~\cite{JSV2011}).

Secondly, it is interesting to investigate what happens when the representation parameter $j$ increases, i.e.\
when the dimension of the representation $W_j$ increases. 
For this purpose, we have plotted the ground state and the first excited state, for a fixed $p$-value, and
for the values $j=10$, $j=30$ and $j=60$ in Figure~3.
These plots remind of the shape of (continuous) paraboson wave functions $\Psi_n^{(a)}(q)$ for increasing values
of $a$. In Figure~4 we have illustrated some of these functions, and the similarity is indeed striking.
This can also be confirmed by a limit calculation. The limit is not simply 
\[
\lim_{j\rightarrow \infty} \phi^{(p)}_{n} (q_{k});
\]
that would just yield zero, as the non-zero contributions are shifted further away from the origin (see Figure~3).
Instead, we need to involve another set of orthogonal polynomials, the dual Hahn polynomials 
$R_n(\lambda(x);\gamma,\delta,N)$ defined by~\cite[(9.6.1)]{Koekoek}
\begin{equation}
R_n(\lambda(x);\gamma,\delta,N) = {\;}_3F_2 \left( \myatop{-n,-x,x+\gamma+\delta+1}{-N,\gamma+1} ; 1 \right), \qquad (n=0,1,\ldots,N),
\label{defR}
\end{equation}
where $\lambda(x)=x(x+\gamma+\delta+1)$. 
These polynomials satisfy a discrete orthogonality relation ($\gamma>-1$, $\delta>-1$):
\begin{equation}
\sum_{x=0}^N \bar w(x;\gamma,\delta,N) R_m(\lambda(x);\gamma,\delta,N)R_n(\lambda(x);\gamma,\delta,N)=
\bar h(n;\gamma,\delta,N)\delta_{mn},
\end{equation}
with $\bar w(x;\gamma,\delta,N)$ and $\bar h(n;\gamma,\delta,N)$ given by~\cite[(9.6.2)]{Koekoek}. Let us
also fix a notation for the orthonormal functions:
\begin{equation}
\tilde R_n(\lambda(x);\gamma,\delta,N)= \sqrt{ \frac{\bar w(x;\gamma,\delta,N)}{\bar h(n;\gamma,\delta,N)} }
 R_n(\lambda(x);\gamma,\delta,N).
\label{Rntilde}
\end{equation}
Consider now some positive parameter $\alpha>0$. 
Let $j$ be the representation parameter, $n=0,1,\ldots,j$, and $0<p<1$.
The following limit is obvious:
\begin{equation}
\lim_{\alpha\rightarrow \infty}  {\;}_3F_2 \left( \myatop{-n,-k,k+2\alpha+1}{-j,2p\alpha+1} ; 1 \right) =
{\;}_2F_1 \left( \myatop{-n,-k}{-j} ; \frac{1}{p} \right).
\end{equation}
In other words,
\begin{equation}
\lim_{\alpha\rightarrow\infty} R_n(\lambda(k);2p\alpha,2(1-p)\alpha,j) = K_n(k;p,j).
\end{equation}
Using some elementary limits for the expressions in~\eqref{Rntilde}, one can in fact show:
\begin{equation}
\lim_{\alpha\rightarrow\infty} \tilde R_n(\lambda(k);2p\alpha,2(1-p)\alpha,j) = \tilde K_n(k;p,j).
\end{equation}
For the wave functions \eqref{phi-even} under consideration, this means
\begin{equation}
\phi^{(p)}_{2n} (q_{k}) = \frac{(-1)^n}{\sqrt{2}} \tilde K_k(n;p,j) = \frac{(-1)^n}{\sqrt{2}} \tilde K_n(k;p,j) = 
\frac{(-1)^n}{\sqrt{2}} \lim_{\alpha\rightarrow\infty}\tilde R_n(\lambda(k);2p\alpha,2(1-p)\alpha,j),
\label{lim-alpha}
\end{equation}
for values $q_k=\sqrt{k}$, $k=1,2,\ldots,j$. 
On the other hand, one finds with~\cite[Eq.~(30)]{JSV2011} that for $jx^2=\lambda(k)=k(k+2\alpha+1)$,
\begin{equation}
\lim_{j\rightarrow\infty} {\;}_3F_2 \left( \myatop{-n,-k,k+2\alpha+1}{-j,2p\alpha+1} ; 1 \right) =
{\;}_1F_1 \left( \myatop{-n}{2p\alpha+1} ; x^2 \right) = \frac{n!}{(2p\alpha+1)_n} L_n^{(2p\alpha)}(x^2),
\label{lim-j1}
\end{equation}
where $L_n^{(a)}$ is a (generalized) Laguerre polynomial.
Following the limit computations in~\cite[\S 4]{JSV2011}, one obtains
\begin{align}
& \lim_{j\rightarrow\infty} \frac{(-1)^n}{\sqrt{2}} j^{1/4} \tilde R_n(j x^2;2p\alpha,2(1-p)\alpha,j) \nonumber\\
& = (-1)^n \sqrt{ \frac{n!}{\Gamma(n+2p\alpha+1)} }|x|^{2p\alpha+1/2} \ee^{-x^2/2} L_n^{(2p\alpha)}(x^2)
= \Psi_{2n}^{(2p\alpha-1)}(x),
\label{lim-j2}
\end{align}
where $\Psi_n^{(a)}(x)$ is the paraboson wave function with parameter $a$ (see~\cite[(A.11)]{JSV2011}). 
Combining~\eqref{lim-alpha} and~\eqref{lim-j2}, it follows indeed that for large $j$-values the
behaviour of the wave functions $\phi_{2n}^{(p)}(q_k)$ is the same as the behaviour of the
paraboson wave functions $\Psi_{2n}^{(2p\alpha-1)}(x)$ for large values of $\alpha$.

For the explanation above, we have used even wave functions; clearly for wave functions of degree $2n+1$
the computation is similar and the conclusion is the same.

\section{The corresponding discrete Fourier transform}

In canonical quantum mechanics, the momentum wave function (in $L^2({\mathbb R})$) is given by the Fourier transform of
the position wave function (and vice versa):
\[
\psi(\bar p)= \frac{1}{\sqrt{2\pi}} \int \ee^{-i\bar p q}\phi(q)dq.
\]
In the present situation we are dealing with discrete wave functions, and an analogue of this should be
viewed as follows. Let 
\begin{equation}
\phi^{(p)} (q_k)=\left(
\begin{array}{c}
\phi_0^{(p)}(q_k)  \\
\phi_1^{(p)} (q_k) \\
\vdots  \\
\phi_{2j}^{(p)}(q_k)
\end{array}
\right), \qquad
\psi^{(p)} (p_k)=\left(
\begin{array}{c}
\psi_0^{(p)}(p_k)  \\
\psi_1^{(p)} (p_k) \\
\vdots  \\
\psi_{2j}^{(p)}(p_k)
\end{array}
\right)\qquad (k=-j,\ldots,+j).
\label{phiV-psiV}
\end{equation}
In this case, the corresponding discrete Fourier transform is defined as the matrix 
$F=(F_{kl})_{-j\leq k,l\leq +j}$ relating these two wave functions:
\begin{equation}
\psi^{(p)}(p_l)=\sum_{k=-j}^j F_{kl}\;\phi^{(p)}(q_k).
\label{F}
\end{equation}
By~\eqref{Uphi}-\eqref{Vpsi}, the columns of $U$ consist of the column vectors $\phi^{(p)}(p_k)$ ($k=-j..,+j$) 
and similarly for the matrix $V$. So~\eqref{F} means that $V = U \cdot F$, or:
\begin{equation}
F=U^T \cdot V = U^T {\cal J} U,
\label{FVU}
\end{equation}
with $J$ given by~\eqref{VJU}.
Using the explicit matrix elements for $U$ and ${\cal J}$,
this leads to the following form of the matrix elements of $F$:
\begin{align}
&F_{j-k,j\mp l}=F_{j+k,j\pm l}=-\frac{\ii}{2} S(k,l;p,j) \pm \frac12 S(k-1,l-1;p,j-1),
\quad (k,l=1,\ldots,j);\label{F1} \\
&F_{j\mp k,j}=F_{j,j\mp k}=-\frac{\ii}{\sqrt{2}} S(k,0;p,j), \quad (k=1,\ldots,j);\label{F2}\\
& F_{jj}=-\ii S(0,0;p,j), \label{F3}
\end{align} 
where
\begin{equation}
S(k,l;p,j) = \sum_{n=0}^j(-1)^n \tilde K_k(n;p,j) \tilde K_l(n;p,j).
\label{S}
\end{equation}
This last expression is easy to simplify, using~\cite[Proposition~3]{Jagan1998} or~\cite[Eq.~(12), p.~85]{Erdelyi}.
One finds:
\begin{equation}
S(k,l;p,j)= \sqrt{\binom{j}{k}\binom{j}{l} } 2^{k+l} (p(1-p))^{(k+l)/2} (1-2p)^{j-k-l}
{\;}_2F_1 \left( \myatop{-k,-l}{-j} ; \frac{1}{4p(1-p)} \right).
\label{S2F1}
\end{equation}
So from \eqref{S2F1} and \eqref{F1}-\eqref{F3} we have an explicit form for the elements $F_{kl}$ of the
matrix $F$.
Just as in~\cite{JSV2011b}, one can state the following properties of the discrete Fourier transform matrix $F$:
\begin{prop}
The $(2j+1)\times(2j+1)$-matrix $F$ is symmetric, $F^T=F$, and unitary, $F^\dagger F=F F^\dagger =I$. 
Furthermore, it satisfies $F^4=I$, so its eigenvalues are $\pm 1, \pm \ii$.
A set of orthonormal eigenvectors of $F$ is given by the rows of $U$, determined in Proposition~\ref{propU}.
The multiplicity of the eigenvalues depends on the parity of $j$. 
When $j=2n$ is even, then the multiplicity of $-\ii,1,\ii,-1$ is $n+1,n,n,n$ respectively.
When $j=2n+1$ is odd, then the multiplicity of $-\ii,1,\ii,-1$ is $n+1,n+1,n+1,n$ respectively.
\end{prop}

\noindent {\bf Proof.}
The symmetry of $F$ is easily seen from the expressions~\eqref{F1}-\eqref{F3}.
The unitarity of $F$ follows from~\eqref{FVU}, the orthogonality of the real matrix $U$ and ${\cal J}^\dagger {\cal J}=I$.
Again using~\eqref{FVU} and the orthogonality of $U$, one finds $F^2=FF^T= V^TUU^TV=V^TV$.
But the explicit form of $V^TV$ is known, see~\eqref{VTV}. Since $(V^TV)^2=I$, the result $F^4=I$ follows. 
So the eigenvalues can only be $\pm 1, \pm \ii$.
From the last part of~\eqref{FVU}, one has $F U^T = U^T {\cal J}$.
So the columns of $U^T$ (or the rows of $U$) form a set of orthonormal eigenvectors
of $F$, and the eigenvalues of $F$ are found in the diagonal matrix ${\cal J}$, see~\eqref{VJU}.  \mybox

We are dealing here with a simple discrete Fourier transform matrix $F$, with parameter $0<p<1$:
the matrix elements of $F$ are given by ${}_2F_1$ expressions~\eqref{S2F1},
the eigenvectors of $F$ are given by the rows of $U$ in Proposition~1,
and the eigenvalues of $F$ (with their multiplicities) are given above.
$F$ still has the property that it transforms position wave functions into momentum wave functions.

Note that due to expression~\eqref{S2F1}, the matrix $F$ is particularly simple when $p=1/2$,
as most of its matrix elements are zero. For $j=3$, the form of $F$ for $p=1/2$ reads
\begin{equation}
2F = \left(\begin{array}{ccccccc}
0&0&1&-\ii\sqrt{2}&-1&0&0\\
0&1&-\ii&0&-\ii&-1&0\\
1&-\ii&0&0&0&-\ii&-1\\
-\ii\sqrt{2}&0&0&0&0&0&-\ii\sqrt{2}\\
-1&-\ii&0&0&0&-\ii&1\\
0&-1&-\ii&0&-\ii&1&0\\
0&0&-1&-\ii\sqrt{2}&1&0&0
\end{array} \right),
\end{equation}
and it is clear how this generalizes for arbitrary $j$.

\section{Conclusions}

We have, for the first time, explored the possibility of using a Lie superalgebra as the basic structure underlying
a finite oscillator model. 
This was inspired by the idea that the position and momentum operators of an oscillator model could most naturally be
represented by odd (rather than even) elements of a superalgebra.
In the case presented here, we have taken the Lie superalgebra $\ssl(2|1)$ for this purpose, being a
simple generalization of the Lie algebra $\su(2)$.

The $\ssl(2|1)$ representations most suitable to use in the model are the $(2j+1)$-dimensional representations $W_j$.
Indeed, in the standard basis for these representations the Hamiltonian is diagonal and the position and momentum
operators are self-adjoint tridiagonal matrices. 
The most general form for the position operator $\hat q$ involves a parameter $p$ ($0<p<1$), see~\eqref{hatq}.
The first main result of the paper is the determination of the eigenvalues and eigenvectors of $\hat q$ in
explicit form (Proposition~1). The spectrum is very simple, see~\eqref{spec-q}; the eigenvectors are in
terms of Krawtchouk polynomials $K_n(x;p,j)$ or $K_n(x;p,j-1)$.
The eigenvalues and eigenvectors of the momentum operator $\hat p$ are similar, see Proposition~2.

The matrix $U$ of eigenvectors of $\hat q$ is interesting from a second point of view. Indeed, its rows correspond to
the discrete position wave functions of the finite oscillator.
These wave functions have been examined in Section~4, by means of plots and by investigating a limit.
Although the discrete wave functions are also given in terms of Krawtchouk polynomials, as in the 
$\su(2)$ case, the behaviour is quite different. 
This is because in the $\su(2)$ case~\cite{Atak2001} the Krawtchouk polynomials appearing in the wave functions are
$K_n(x;\frac12,2j+1)$, whereas here we have a combination of $K_n(x;p,j)$ and $K_n(x;p,j-1)$.
For large values of $j$, the discrete wave functions tend to certain paraboson wave functions.

Since the position and momentum wave functions have fairly simple forms, we are able to construct the matrix $F$
that transforms them into each other. 
This matrix is a discrete analogue of the Fourier transform.
The matrix elements of $F$ are terminating ${\,}_2F_1$ series. 
The discrete version of the Fourier transform $F$ has many properties in common with the standard Discrete Fourier Transform,
see Proposition~3.

Apart from the finite-dimensional representations $W_j$ considered here, the Lie superalgebra $\ssl(2|1)$ also has
an interesting class of infinite-dimensional representations in $\ell^2({\mathbb Z}_+)$.
It would be tempting to study $\ssl(2|1)$ oscillator models in these representations, 
and we hope to tackle this problem in a future paper.

\section*{Acknowledgments}
E.I.~Jafarov was supported by a postdoc fellowship from the Azerbaijan National Academy of Sciences.

\newpage
\begin{figure}[th]
\[
\begin{tabular}{lc}
\hline\\[-3mm]
(a)& \includegraphics[scale=1]{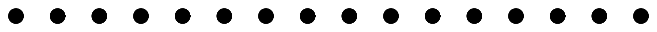}\\[2mm]
(b)& \includegraphics[scale=1]{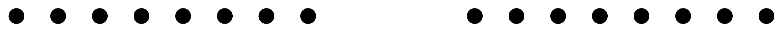}\\[2mm]
(c)& \includegraphics[scale=1]{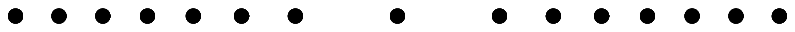}\\[2mm]
(d)& \includegraphics[scale=1]{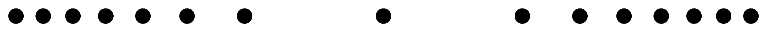}\\[2mm]
\hline
\end{tabular} 
\]
\caption{Plots of a typical spectrum of the position operator: (a) for the $\su(2)$ model~\cite{Atak2001};
(b) for the $\uu(2)_\alpha$ model~\cite{JSV2011};
(c) for the $\su(2)_\alpha$ model~\cite{JSV2011b};
(d) for the $\ssl(2|1)$ model.}
\label{fig1}
\end{figure}

\newpage
\begin{figure}[th]
\begin{tabular}{ccc}
\hline\\[-3mm]
\includegraphics[scale=0.50]{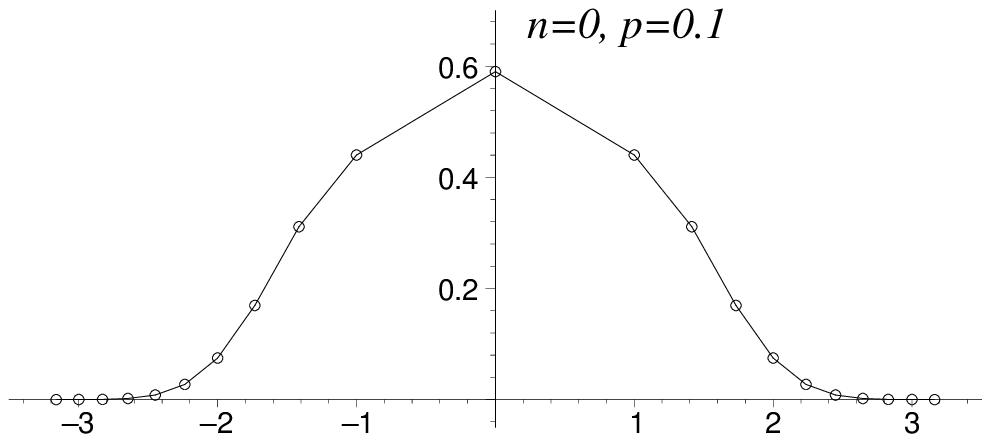} & \includegraphics[scale=0.50]{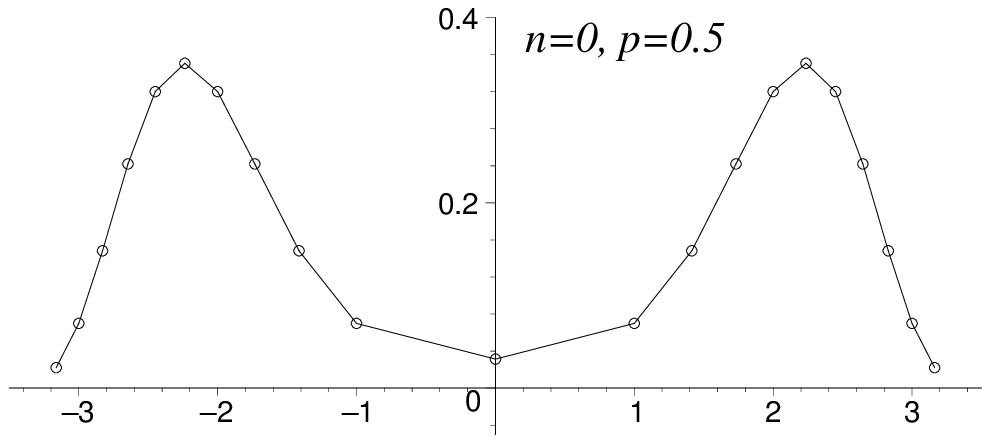} & \includegraphics[scale=0.50]{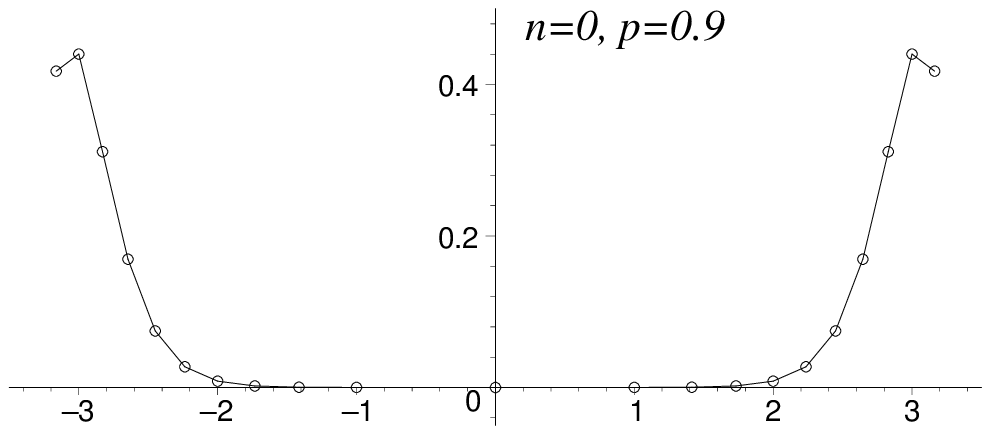}\\[2mm]
\includegraphics[scale=0.50]{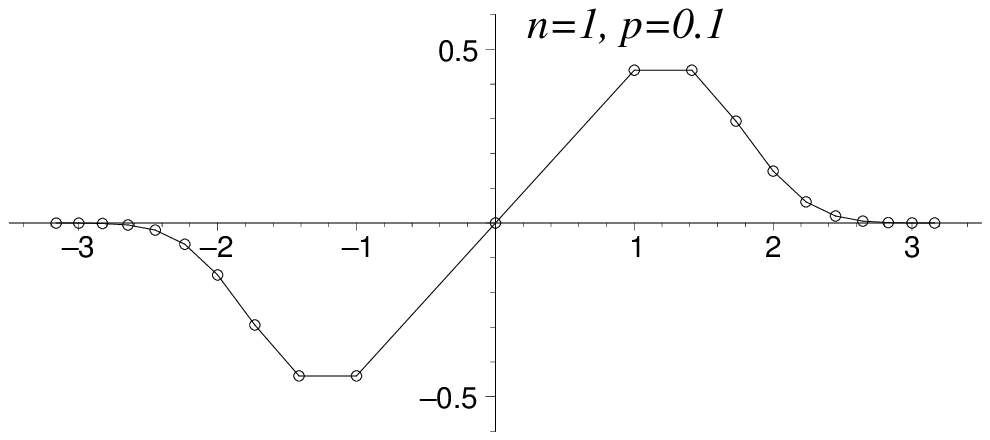} & \includegraphics[scale=0.50]{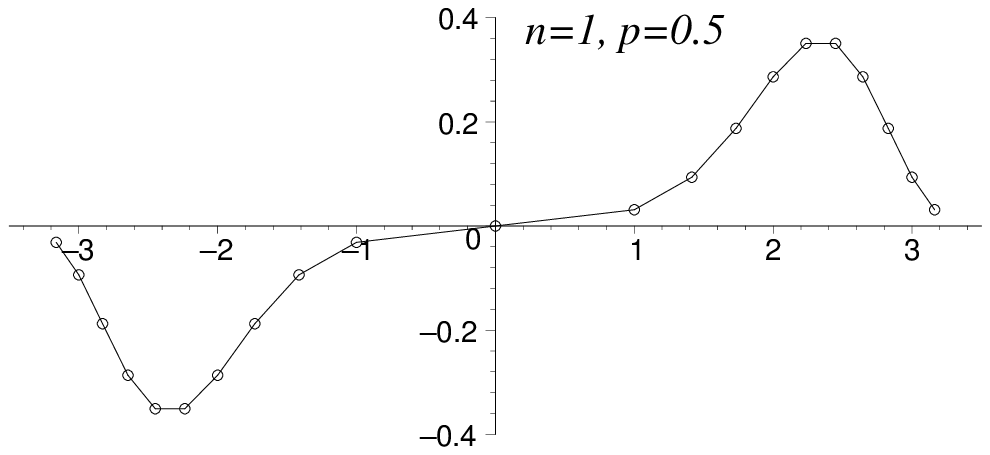} & \includegraphics[scale=0.50]{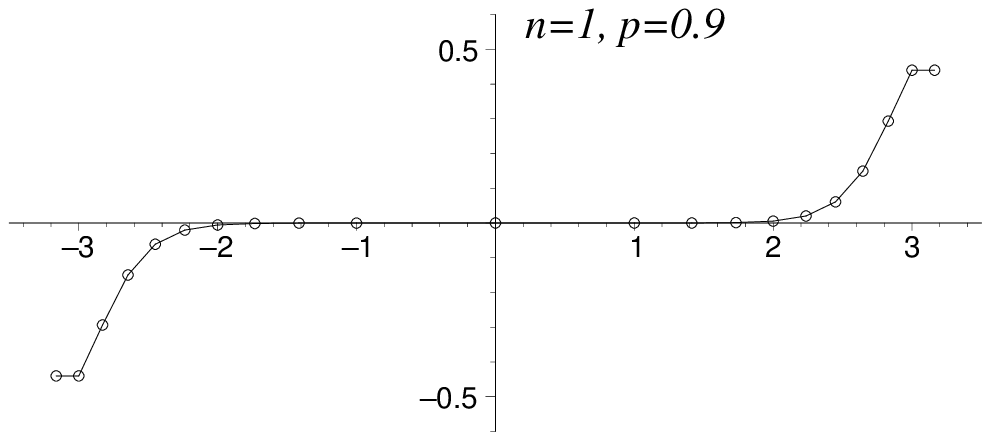}\\[2mm]
\includegraphics[scale=0.50]{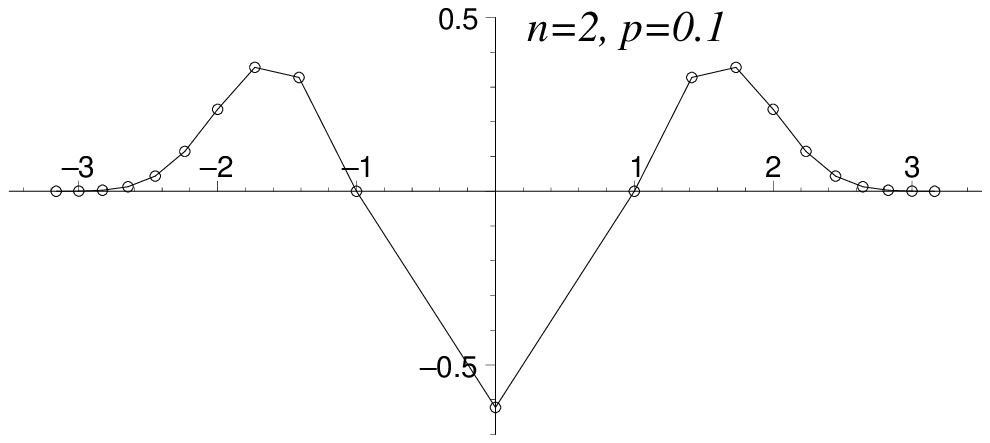} & \includegraphics[scale=0.50]{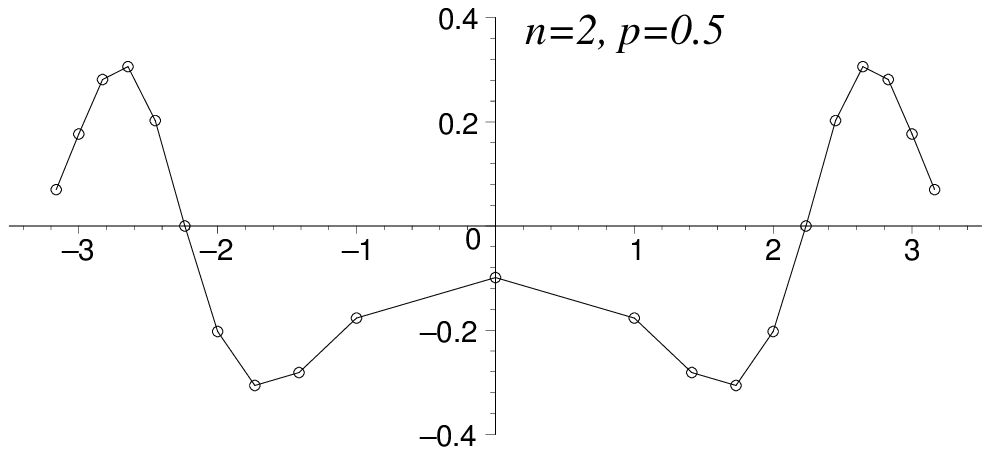} & \includegraphics[scale=0.50]{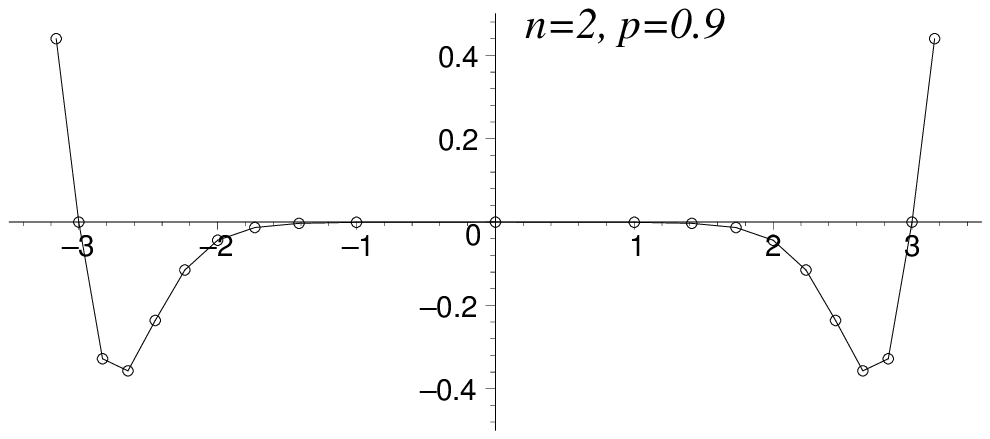}\\[2mm]
\includegraphics[scale=0.50]{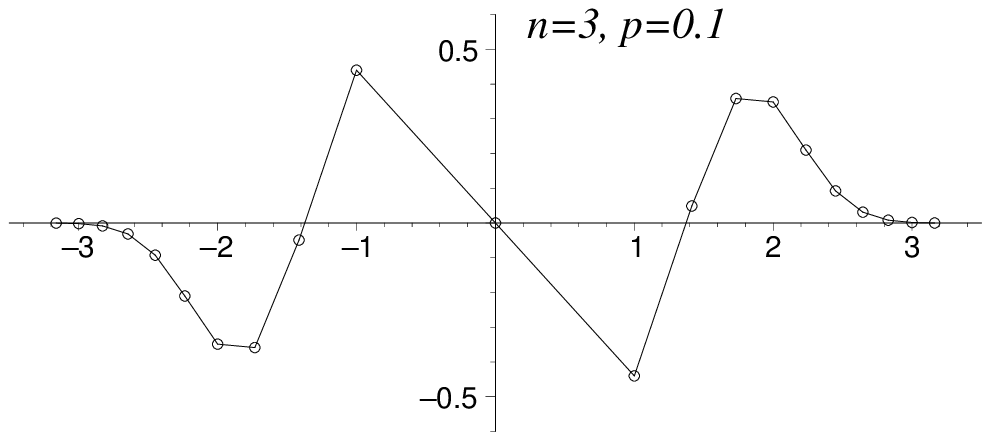} & \includegraphics[scale=0.50]{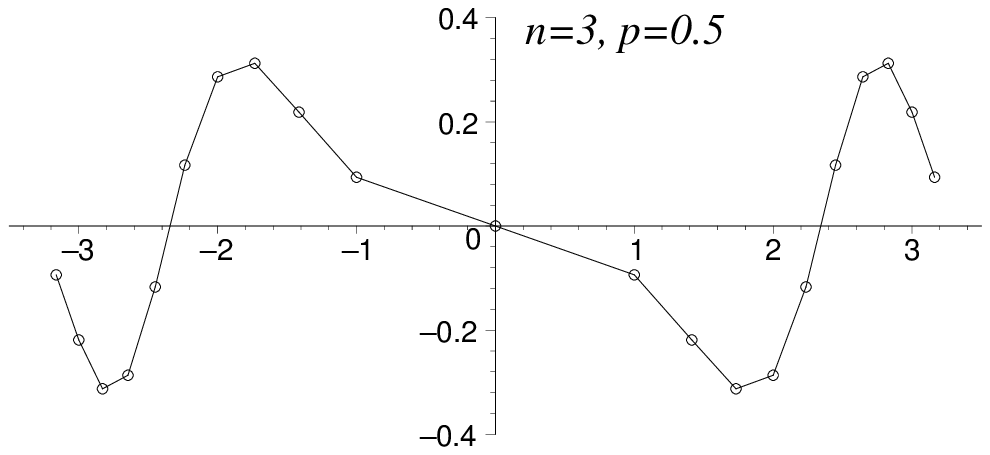} & \includegraphics[scale=0.50]{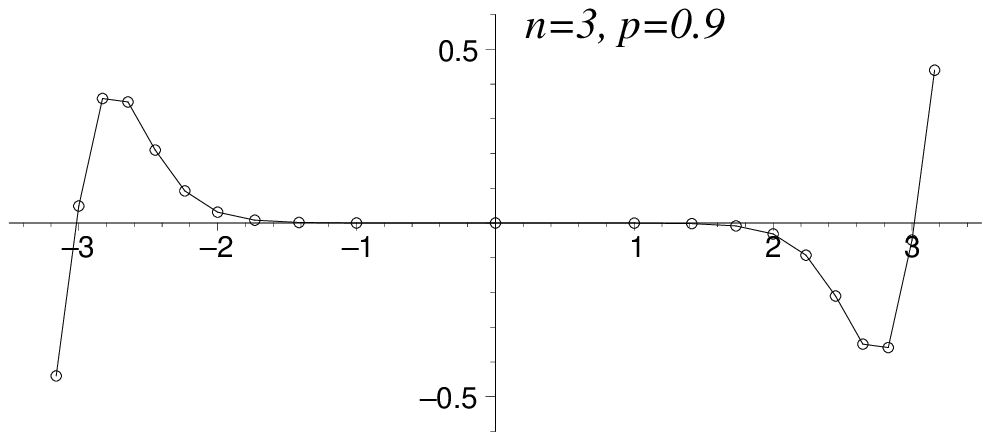}\\
\hline
\end{tabular} 
\caption{Plots of the discrete wave functions $\phi^{(p)}_n(q)$ in the representation with $j=10$,
for $n=0,1,2,3$ and for $p=0.1$ (left column), $p=0.5$ (middle column) and $p=0.9$ (right column).}
\label{fig2}
\end{figure}

\newpage
\begin{figure}[th]
\[
\begin{tabular}{ccc}
\hline\\[-3mm]
\includegraphics[scale=0.60]{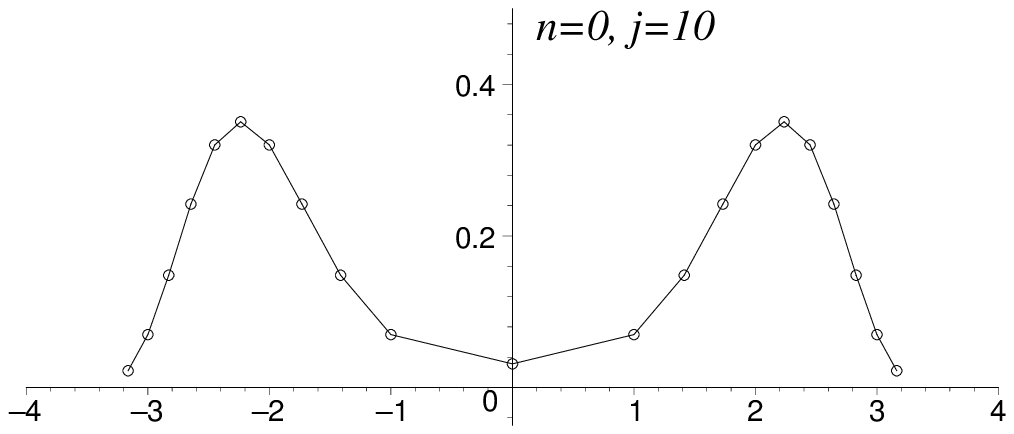} & & \includegraphics[scale=0.60]{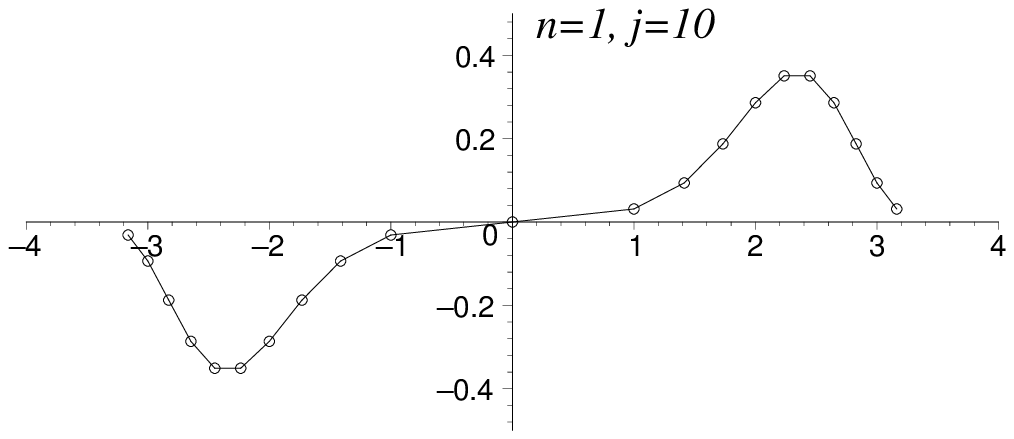}\\[2mm]
\includegraphics[scale=0.60]{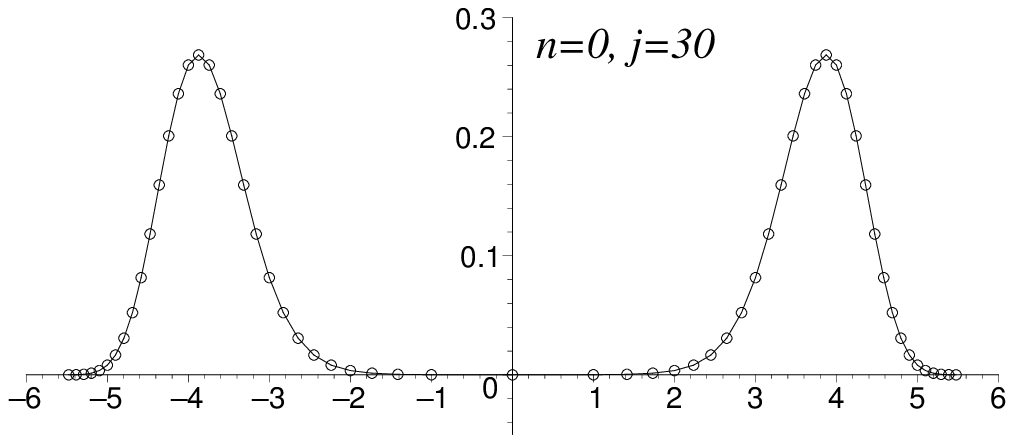} & &\includegraphics[scale=0.60]{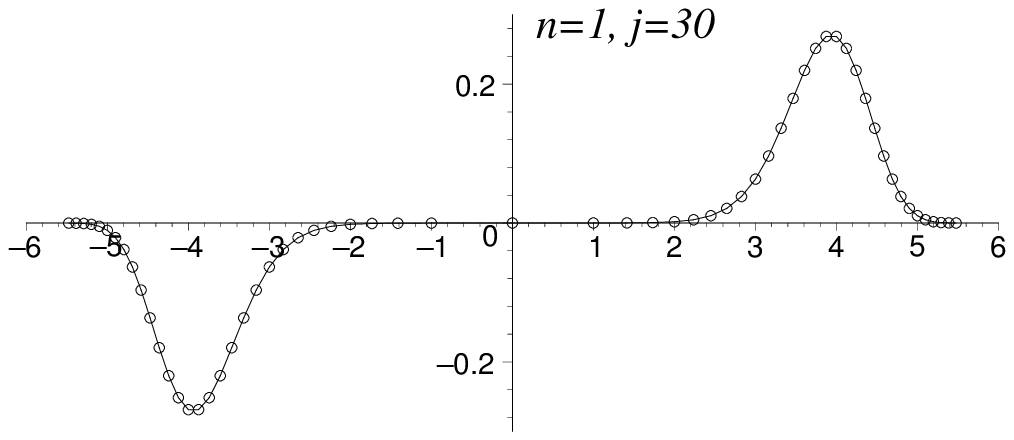}\\[2mm]
\includegraphics[scale=0.60]{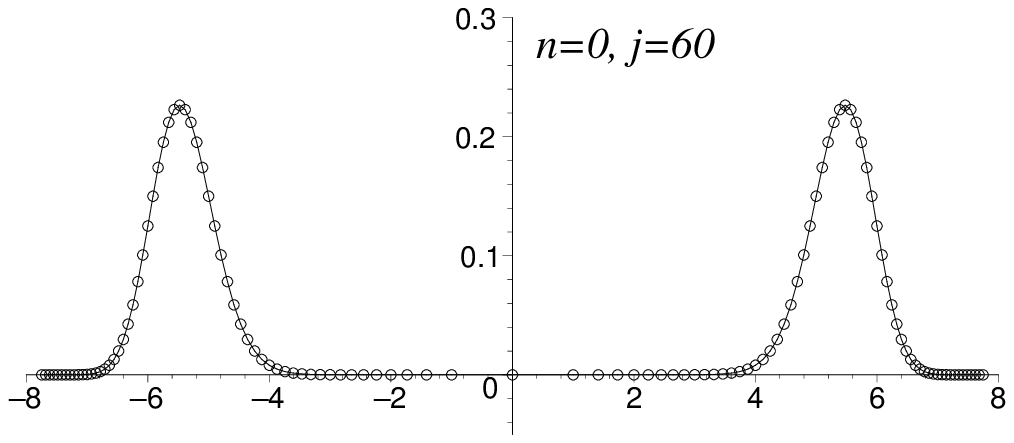} & & \includegraphics[scale=0.60]{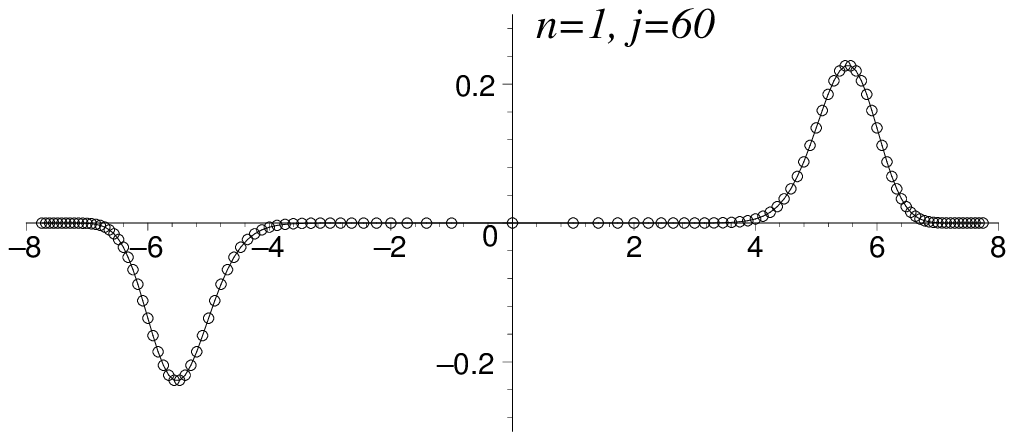}\\
\hline
\end{tabular} 
\]
\caption{Plots of the discrete wave functions $\phi^{(p)}_n(q)$ for fixed $p=0.5$, for $n=0$ (left column)
and $n=1$ (right column), but with $j$ varying: $j=10,30,60$.}
\label{fig3}
\end{figure}

\newpage
\begin{figure}[th]
\[
\begin{tabular}{ccc}
\hline\\[-3mm]
\includegraphics[scale=0.60]{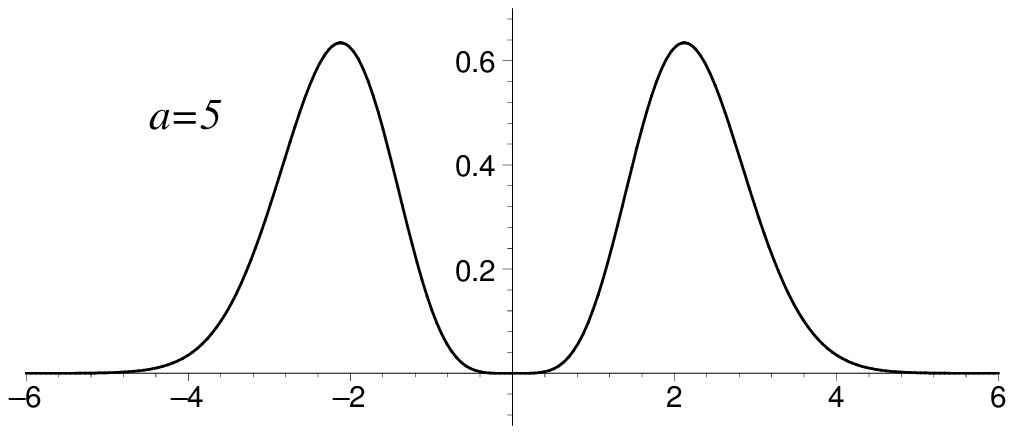} && \includegraphics[scale=0.60]{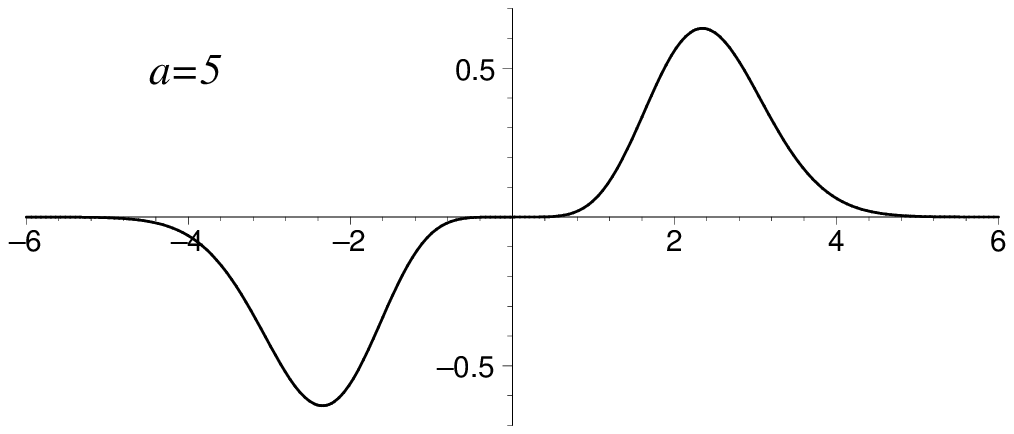}\\[2mm]
\includegraphics[scale=0.60]{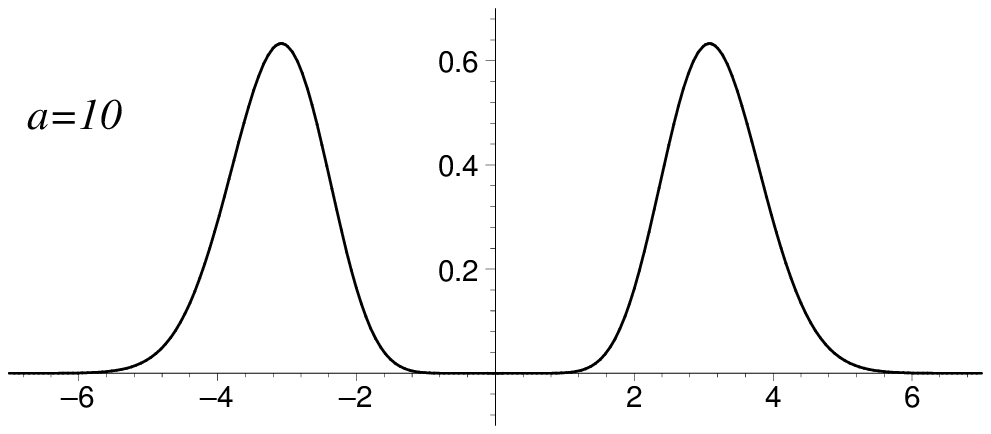} && \includegraphics[scale=0.60]{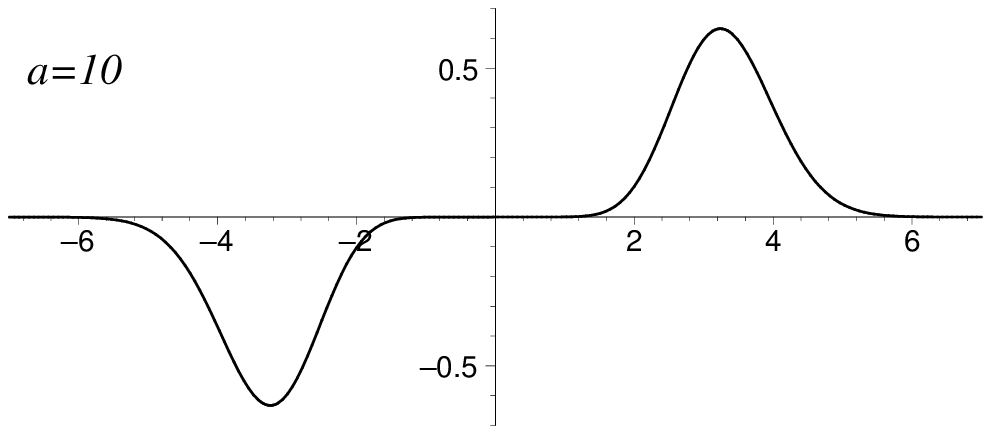}\\[2mm]
\includegraphics[scale=0.60]{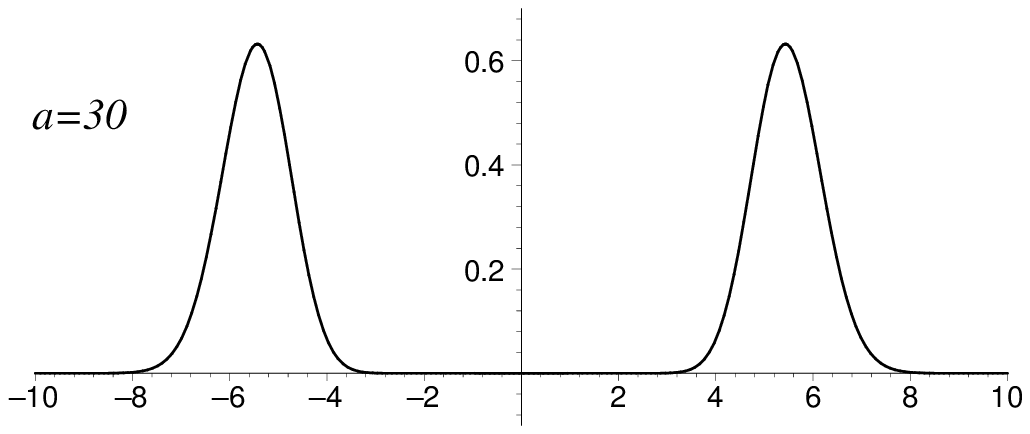} && \includegraphics[scale=0.60]{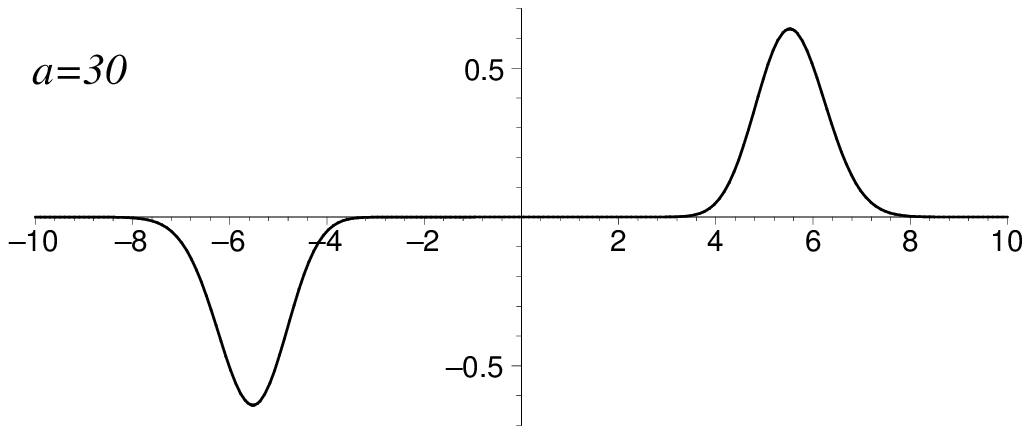}\\
\hline
\end{tabular} 
\]
\caption{Plots of the continuous paraboson wave functions $\Psi^{(a)}_n(x)$ for $n=0$ (left column)
and $n=1$ (right column), and with $a$ varying: $a=5,10,30$.}
\label{fig4}
\end{figure}


\begin{thebibliography}{99}

\bibitem{Andrews}
G.E.\ Andrews, R.\ Askey and R.\ Roy,
{\em Special functions}
(Cambridge University Press, Cambridge, 1999).

\bibitem{Atak1994}
N.M.\ Atakishiyev and K.B.\ Wolf,
%Approximation on a finite set of points through Kravtchuk functions,
Rev.\ Mex.\ Fis. {\bf  40}, 366-377 (1994).

\bibitem{Atak1997}
N.M.\ Atakishiyev and K.B.\ Wolf,
%Fractional Fourier-Kravtchuk transform,
J.\ Opt.\ Soc.\ Am.\ A. {\bf  14}, 1467-1477 (1997).

\bibitem{Atak1999b}
N.M.\ Atakishiyev, L.E.\ Vicent and K.B.\ Wolf,
%Continuous versus Discrete Fractional Fourier Transforms,
J.\ Comp.\ Appl.\ Math. {\bf  107}, 73-95 (1999).

\bibitem{Atak2001}
N.M.\ Atakishiyev, G.S.\ Pogosyan, L.E.\ Vicent and K.B.\ Wolf,
%Finite two-dimensional oscillator: I. The cartesian model,
J.\ Phys.\ A {\bf 34}, 9381-9398 (2001).

\bibitem{Atak2001b}
N.M.\ Atakishiyev, G.S.\ Pogosyan, L.E.\ Vicent and K.B.\ Wolf,
%Finite two-dimensional oscillator: I. The radial model,
J.\ Phys.\ A {\bf 34}, 9399-9415 (2001).

\bibitem{Atak2005}
N.M.\ Atakishiyev, G.S.\ Pogosyan and K.B.\ Wolf,
%Finite models of the oscillator,
Phys.\ Part.\ Nuclei {\bf  36}, 247-265 (2005).

\bibitem{Bailey}
W.N.\ Bailey,
{\em Generalized hypergeometric series} 
(Cambridge University Press, Cambridge, 1964).

\bibitem{Erdelyi}
A.\ Erd\'elyi, W.\ Magnus, F.\ Oberhettinger, F.G.\ Tricomi, 
{\em Higher Transcendental Functions}, Volume~1
(McGraw-Hill, New York, 1953).

\bibitem{Frappat}
L.\ Frappat, A.\ Sciarrino and P.\ Sorba,
{\em Dictionary on Lie Algebras and Superalgebras}
(Academic Press, London, 2000).

\bibitem{Ismail}
M.E.H.\ Ismail,
{\em Classical and quantum orthogonal polynomials in one variable}
(Cambridge University Press, Cambridge, 2005).

\bibitem{JLV2008}
E.\ Jafarov, S.\ Lievens and J.\ Van der Jeugt,
%The Wigner distribution function for the one-dimensional parabose oscillator.
J.\ Phys.\ A {\bf 41}, 235301 (2008).

\bibitem{JSV2011}
E.I.\ Jafarov, N.I.\ Stoilova and J.\ Van der Jeugt,
%Finite oscillator models: the Hahn oscillator,
J.\ Phys.\ A {\bf 44}, 265203 (2011).

\bibitem{JSV2011b}
E.I.\ Jafarov, N.I.\ Stoilova and J.\ Van der Jeugt,
%The su(2)a Hahn oscillator and a discrete Fourier-Hahn transform,
J.\ Phys.\ A {\bf 44}, 355205 (2011).


\bibitem{Koekoek}
R.\ Koekoek, P.A.\ Lesky and R.F.\ Swarttouw,
{\em Hypergeometric orthogonal polynomials and their $q$-analogues} 
(Springer-Verlag, Berlin, 2010).

\bibitem{Marcu}
M.\ Marcu,
%The representations of spl(2|1): an example of representations of basic superalgebras,
J.\ Math.\ Phys. {\bf 21}, 1277-1283 (1980).

\bibitem{Mukunda}
N.\ Mukunda, E.C.G.\ Sudarshan, J.K.\ Sharma and C.L.\ Mehta,
%REPRESENTATIONS AND PROPERTIES OF PARA-BOSE OSCILLATOR OPERATORS .1. ENERGY POSITION AND MOMENTUM EIGENSTATES,
J.\ Math.\ Phys. {\bf 21}, 2386-2394 (1980).

\bibitem{Ohnuki}
Y.\ Ohnuki and S.\ Kamefuchi,
{\em Quantum Field Theory and Parastatistics}
(Springer-Verlag, New-York, 1982).

\bibitem{Palev79}
T.D.\ Palev,
%  Lie-superalgebraical approach to the second quantization
Czech J.\ Phys., Sect. {\bf B29}, 91-98 (1979).

\bibitem{Palev82}
T.D.\ Palev,
% Wigner approach to quantization. Noncanonical quantization of two particles interacting via a harmonic potential 
J.\ Math.\ Phys. {\bf 23}, 1778-1784 (1982).

\bibitem{Scheunert1977}
M.\ Scheunert, W.\ Nahm and V.\ Rittenberg,
% Irreducible representations of the osp(1,2) and spl(1,2) graded Lie algebras,
J.\ Math.\ Phys. {\bf 18}, 155-162 (1977).

\bibitem{Shiri2006}
M.\ Shiri-Garakani and D.\ Finkelstein,
%Finite quantum kinematics of the harmonic oscillator,
J.\ Math.\ Phys. {\bf 47}, 032105 (2006).

\bibitem{Slater}
L.J.\ Slater,
{\em Generalized hypergeometric functions} 
(Cambridge University Press, Cambridge, 1966).

\bibitem{Jagan1998}
J.\ Van der Jeugt and R.\ Jagannathan,
%Realizations of su(1,1) and U_q(su(1,1)) and generating functions for orthogonal polynomials,
J.\ Math.\ Phys. {\bf 39}, 5062-5078 (1998). 

\bibitem{Wigner}
E. P.\ Wigner, 
% Do the equations of motion determine the quantum mechanical commutation relations? 
Phys.\ Rev. {\bf 77}, 711-712 (1950).


\end{thebibliography}
\end{document}